\documentclass[aps,prd,reprint,superscriptaddress,floatfix,nofootinbib]{revtex4-2}

\usepackage{mathtools,bm}
\usepackage{graphicx}
\usepackage{hyperref}
\usepackage{enumerate}
\usepackage[dvipsnames]{xcolor} 
\usepackage{physics}
\usepackage{braket}

\newcommand{\id}{\hat\openone}
\renewcommand{\a}[1]{\hat{a}_{#1}}
\newcommand{\ad}[1]{\hat{a}_{#1}^\dagger}

\newcommand{\ii}{\mathrm{i}}
\newcommand{\diff}[1]{\mathrm{d}#1\,}
\renewcommand*\d[2]{
	\mathrm{d}
	\ifx\relax#1\relax\else
	\rule{-0.02em}{1.5ex}^{#1}\rule{0.08em}{0ex}\!
	\fi
	#2\,
}
\DeclareMathOperator{\sign}{sgn}
\DeclareMathOperator{\Heaviside}{\Theta}

\begin{document}
\title{Derivative coupling enables genuine entanglement harvesting in causal communication 
}

\author{Adam Teixid\'{o}-Bonfill}
\email{adam.teixido-bonfill@uwaterloo.ca}
\affiliation{Department of Applied Mathematics, University of Waterloo, Waterloo, Ontario, N2L 3G1, Canada}
\affiliation{Institute for Quantum Computing, University of Waterloo, Waterloo, Ontario, N2L 3G1, Canada}
\affiliation{Perimeter Institute for Theoretical Physics, Waterloo, Ontario, N2L 2Y5, Canada}

\author{Eduardo Mart\'{i}n-Mart\'{i}nez}
\email{emartinmartinez@uwaterloo.ca}

\affiliation{Department of Applied Mathematics, University of Waterloo, Waterloo, Ontario, N2L 3G1, Canada}
\affiliation{Institute for Quantum Computing, University of Waterloo, Waterloo, Ontario, N2L 3G1, Canada}
\affiliation{Perimeter Institute for Theoretical Physics, Waterloo, Ontario, N2L 2Y5, Canada}

\begin{abstract}
We show that particle detectors coupled to a massless quantum field through its derivative can genuinely harvest entanglement form the field even when they are in causal contact in flat spacetime. This is particularly relevant since the derivative coupling model captures some interesting experimentally realizable systems and since the harvested entanglement peaks at full light contact.\\ 
\end{abstract}

\maketitle
\section{Introduction}

It is well known that physical states of a quantum field (such as e.g., the Minkowski vacuum) contain entanglement between the degrees of freedom of different spacetime regions, including regions that are spacelike separated~\cite{Summers1985TheVV,Summers1987}. In fact, Bell inequalities can be violated by performing measurements on the local algebras of observables of spatially separated regions \cite{Reznik2005}. In later years, a protocol known as \textit{entanglement harvesting} was developed through which that entanglement can be transferred to local probes. The protocol consists of a pair of two initially uncorrelated particle detectors which interact locally with a quantum field in a way that the detectors end up entangled through the extraction of the entanglement contained in the field state~\cite{Valentini1991,Reznik2003,Pozas2015}. When the two detectors interact with the field in spacelike separated regions, it is clear that any entanglement that the detectors acquire has to be transferred from pre-existing entanglement in the field, as the detectors cannot communicate (see, e.g.,~\cite{cavitySignaling2014,infoWOenergy2015,detectorsSignaling2015,Pipo2023Signaling}). In \cite{Erickson2021_When}, it was pointed out that, when the detectors are causally connected, the detectors can get entangled through their communication through the field as well as through genuine harvesting depending on the details of the particular configurations of the detectors.

Previous literature focused on clarifying this distinction in the context of the usual Unruh-DeWitt detector model~\cite{Unruh1976,DeWitt,Unruh-Wald}, where the coupling between the detectors and the field is prescribed to be a linear coupling to the field amplitude. In previous work it was found that for massless fields in 1+1 and 3+1 dimensional flat spacetime, while entanglement between the two detectors peaks when the detectors are light connected, the entanglement that the two detectors acquire is not coming from harvesting from the field and it is rather due to the fact that the detectors communicate \cite{Erickson2021_When}. This previous work also found that genuine harvesting in flat spacetime can  actually contribute to acquired entanglement even when the detectors are light connected. However, this required either the space dimension to be $n\geq5$ or using a massive field with enough mass. It was found later on that in the curved spacetimes where there can be caustics and secondary null geodesics connecting the trajectories of the two detectors, it is possible to find regimes where for light-connected detectors where the entanglement they  acquire is due to genuine harvesting~\cite{Lensing2023}. In this paper we will see that, even for massless fields in 1+1D and 3+1D Minkowski spacetime, it is possible to build setups where genuine harvesting can dominate even when the detectors are in their strongest possible causal contact, and therefore maximize the amount of entanglement (genuinely) harvested by two detectors while they are able to communicate. 

To observe this behavior, unlike in the most common approaches to entanglement harvesting, here we will prescribe the interaction as the detector coupling to the derivative of the field amplitude and discuss the main differences with the much more common amplitude coupling. 

Models of particle detectors that are coupled to the (proper) time derivative of the field amplitude have been primarily used to sidestep IR divergences that appear at low dimensions for massless fields \cite{Raine1991,Raval1996,Eduardo2014Zero,Juárez-Aubry_2014,Wang2014,Juárez-Aubry2018,Tjoa2020,Juárez-Aubry2022,DBunney_2023}. 
While these IR divergences can be regularized by a cutoff, the final density matrix of the detectors do depend (albeit weakly) on the chosen cutoff, see, e.g., \cite{Pozas2015,Eduardo2014Zero,Juárez-Aubry_2014}. Coupling the detectors to the derivative of the field naturally removes this IR divergence. Additionally, the 1+1D derivative coupling short distance behaviour resembles that of the 3+1D amplitude coupling~\cite{Juárez-Aubry_2014,Juárez-Aubry2018}, and dualities between derivative and amplitude coupling were found in \cite{Matheus2023Duality}. These facts have made the derivative coupling an interesting way to model some of the behaviour of particle detectors coupling to amplitude in 3+1 dimensions by using a simpler model in 1+1 dimensions.

Remarkably, the derivative coupling connects with very physical and experimentally accessible models. The coupling to the derivative of a scalar field is a better mimicker of some aspects of the interaction of an atom with the electromagnetic field \cite{Pozas2016Electromagnetic,Richard2021LightMatter}. Furthermore, particle detectors that couple to the field derivative can indeed be realized experimentally in superconducting circuits, as shown, for example in \cite{Emma2017,Tunable2023}. In these systems, the particle detector is a superconducting qubit coupled to a transmission line, which serves as a 1+1D massless field. The interaction between the qubit and the transmission line is typically modelled with the spin-boson model~\cite{Spin-Boson1987}, with an Ohmic bosonic bath, see, e.g., \cite{Peropadre2013,Forn-Diaz2017,Tunable2023}. This description turns out to be none other than the derivative coupling in the language of particle detectors~\cite{Emma2017}.

The article is organized as follows: Section \ref{sec:protocol} describes an entanglement harvesting setup with derivative coupling. Section \ref{sec:communication} reviews how to separate the gathered entanglement into communication and genuine entanglement harvesting. Section \ref{sec:results} compares the communication contributions for amplitude and derivative coupling, and provides examples where derivative coupling allows to genuinely harvest entanglement while in full causal communication. In this paper, we use natural units $\hbar=c=1$, and indicate spacetime points (represented by their coordinates in an inertia frame) as $\mathsf{x}\equiv(t,\bm{x})$.

\section{Entanglement harvesting protocol}
\label{sec:protocol}

Here we provide a summary of the entanglement harvesting protocol, using particle detectors that couple to the time derivative of the field amplitude. 

Consider a real massless scalar quantum field on (\mbox{$n$+1})-dimensional Minkowski spacetime. If we fix some inertial frame defining coordinates $(t,\bm x)$, the field amplitude can be expanded in plane-wave modes as
\begin{equation}
    \hat \phi(t,\bm{x}) = \int \frac{\d{n}{\bm{k}}}{\sqrt{2 (2\pi)^n \omega_{\bm{k}}}} (e^{\ii (\omega_k t-\bm{k}\cdot\bm{x})} \ad{\bm{k}}  + \text{H.c.}),
\end{equation}
where $\omega_{\bm{k}}=|\bm{k}|$ for a massless field, and the creation and annihilation operators obey $[\a{\bm{k}},\ad{\bm{k}'}]=\delta^n(\bm{k}-\bm{k}')\hat\openone$. 

Consider two inertial detectors, A and B, comoving with the frame $(t,\bm x)$. Let the detectors be two-level quantum systems that interact locally with the quantum field. Each detector has a free Hamiltonian whose eigenstates are labeled as the ground state $\ket{g_\nu}$  and the excited state $\ket{e_\nu}$, with $\nu\in\{\text{A},\text{B}\}$. The free Hamiltonian of each detector sets an energy gap $\Omega_\nu$ between the two detectors. The detectors couple to the derivative of the field amplitude in accordance to the following Unruh-DeWitt interaction Hamiltonian (in the interaction picture)~\cite{Unruh1976,DeWitt,Unruh-Wald
}:
\begin{align}
    &\hat{H}_I(t)=\hat{H}_\textsc{a}(t) + \hat{H}_\textsc{b}(t),\nonumber\\
    &\hat{H}_\nu(t) = \lambda_\nu \chi_\nu(t) \hat\mu_\nu(t) \int \d{n}{\bm{x}}F_\nu(\bm{x})\partial_t\hat\phi(t,\bm{x}).
\end{align}

The interaction strength is controlled by $\lambda_\nu$, $\chi_\nu(t)$ and $F_\nu(\bm{x})$ respectively are the switching function and the smearing function, which localize the interaction in time and space. The monopole moment of each detector is given by
\begin{equation}
    \hat\mu_\nu(t) = e^{\ii \Omega_\nu t}\hat\sigma^+_\nu +  e^{-\ii \Omega_\nu t}\hat\sigma^-_\nu,
\end{equation}
where $\hat\sigma^+_\nu=\ketbra{e_\nu}{g_\nu}=(\hat\sigma^-_\nu)^\dagger$. 

\begin{figure}[ht]
    \centering
    \includegraphics[width=0.75\columnwidth]{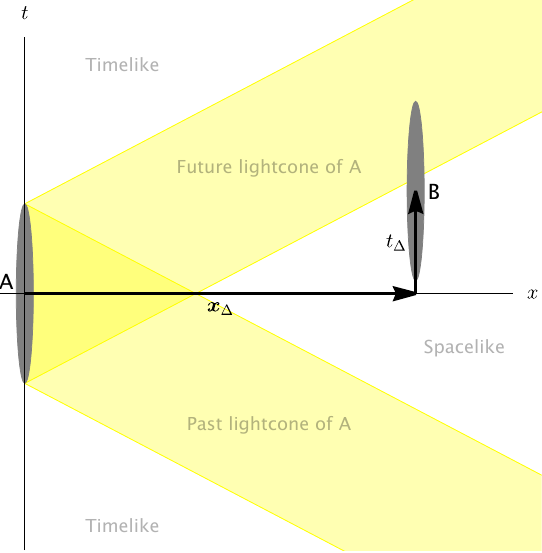}
    \caption{Spacetime diagram depicting with grey ellipses the regions where detectors A and B interact (strongly) with the field. The diagram shows the delay $t_\Delta$ and separation $\bm{x}_\Delta$, and the yellow shaded regions indicate null separation from the detector A.}
    \label{fig:HarvestingDiagram}
\end{figure}

For simplicity, we take the detectors to be equal, with $\lambda =\lambda_\nu$ and $\Omega=\Omega_\nu$. Moreover, we use the same switching and smearing functions for both detectors, (up to a spacetime translation)
\begin{equation}
    \chi_\nu(t)=\chi(t-t_\nu),\ F_\nu(\bm{x})=F(\bm{x}-\bm{x}_\nu).\label{eq:translation}
\end{equation}
The delay between switchings $t_\Delta=t_\textsc{b}-t_\textsc{a}$ and the separation $\bm{x}_\Delta=\bm{x}_\textsc{b}-\bm{x}_\textsc{a}$ control the causal relations between the localized interactions, as depicted in Figure~\ref{fig:HarvestingDiagram}.

The evolution of the initial state of the full system initially in a state $\hat \rho_0$ yields the evolved state $\hat \rho =\hat U \hat\rho_0 \hat U^\dagger$. We can evaluate the time evolution operator $\hat U$ through its Dyson series:
\begin{align}
    \hat U =& \underbrace{\!\!\!\!\!\!\phantom{\int}\hat\openone}_{\hat U^{(0)}} \underbrace{- \ii\int\!\diff {t} \hat H_I(t)}_{\hat U^{(1)}} 
    \underbrace{-\int\!\diff {t}\! \int^t\! \diff {t'} \hat H_I(t)\hat H_I(t')}_{\hat U^{(2)}}+ \mathcal{O}(\lambda^3). \label{eq:DysonSeries}
\end{align}
Denoting $\hat \rho^{(i,j)}=\hat U^{(i)}\hat\rho_0\hat U^{(j)\dagger}$, the final state becomes
\begin{equation}
    \hat \rho = \hat\rho_0+\hat \rho^{(1,0)}+\hat \rho^{(0,1)}+\hat \rho^{(2,0)}+\hat \rho^{(1,1)}+\hat \rho^{(0,2)}+\mathcal{O}(\lambda^3).\label{eq:finalStateSeries}
\end{equation}
To study the entanglement the detectors acquire through the interaction, we need to start in an initially uncorrelated detector state, and after the interaction trace out the field, yielding  \mbox{$\hat \rho_\textsc{ab} = \Tr_{\phi}(\hat\rho)$}, and then look at the correlations in the evolved state $\hat \rho_\textsc{ab}$. Again to take a simple case, let us take the common starting assumption that the detectors and the field start uncorrelated and in their ground states,
\begin{equation}
    \hat\rho_0 = \ketbra{g_\textsc{a}}{g_\textsc{a}}\otimes\ketbra{g_\textsc{b}}{g_\textsc{b}}\otimes\ketbra{0}{0}.
\end{equation}
where $\ket{0}$ is the Minkowski vacuum. Due to the choice of initial state, all the contributions to $\hat \rho_\textsc{ab}$ which are of odd order in $\lambda$ vanish.

Under the given assumptions, the final state of the detectors to leading order in the coupling strength, represented in matrix form in the basis $\{\ket{g_\textsc{a}g_\textsc{b}}$, $\ket{e_\textsc{a}g_\textsc{b}}$, $\ket{g_\textsc{a}e_\textsc{b}}$, $\ket{e_\textsc{a}e_\textsc{b}}\}$ is
\begin{equation}
    \hat\rho_\textsc{ab} = \begin{pmatrix}
    1-2\mathcal L & 0 & 0 & \mathcal{M}^*\\
    0 & \mathcal L & \mathcal L_{\textsc{a}\textsc{b}} & 0 \\
    0 & \mathcal {L}_{\textsc{a}\textsc{b}}^* & \mathcal L & 0 \\
    \mathcal{M} & 0 & 0 & 0 \\
    \end{pmatrix} + \mathcal{O}(\lambda^4),\label{eq:rho_qubits}
\end{equation}
where the local noise is
\begin{align}
    &\mathcal L= \lambda^2\int\diff{t} \diff{t'}\d{n}{\bm{x}}\d{n}{\bm{x}'}  \chi(t)\chi(t')F(\bm x)F(\bm x')\nonumber\\
    &\qquad\qquad \cross e^{\ii\Omega\Delta t}W_{tt'}^*(\Delta t,\Delta\bm{x}),
\end{align}
and
\begin{align}
    &\mathcal L_\textsc{ab}= \lambda^2e^{-\ii\Omega t_\Delta} \int \diff{t} \diff{t'} \d{n}{\bm{x}}\d{n}{\bm{x}'} \chi(t)\chi(t')F(\bm x)F(\bm x')\nonumber\\
    &\qquad\qquad \cross e^{\ii\Omega\Delta t}W_{tt'}^*(\Delta t-t_\Delta,\Delta\bm{x}-\bm{x}_\Delta),
    \nonumber \\
    &\mathcal M=-\lambda^2e^{\ii\Omega(t_\textsc{a}+t_\textsc{b})}\int \diff{t} \diff{t'}\d{n}{\bm{x}}\d{n}{\bm{x}'} \chi(t)\chi(t')F(\bm x)F(\bm x')\nonumber\\
    &\qquad\qquad \cross e^{\ii\Omega(t+t')}W_{tt'}(|\Delta t-t_\Delta|,\Delta\bm{x}-\bm{x}_\Delta),\label{eq:LMSimple}
\end{align}
where $\Delta t = t-t'$ and $\Delta\bm{x}=\bm{x}-\bm{x}'$, $t_\Delta=t_\textsc{b}-t_\textsc{a}$, $\bm{x}_\Delta=\bm{x}_\textsc{b}-\bm{x}_\textsc{a}$, and $W_{tt'} = \partial_t\partial_{t'} W$, where $W$ is the Wightman function of the vacuum of ($n$+1)-dimensional Minkowski spacetime,\footnote{Notice that in the more common case where the detectors couple to the amplitude instead of the derivative, the expressions in Eq.~\eqref{eq:LMSimple} remain the same except for replacing $W_{tt'}\to W$. }
\begin{align}
    W(\Delta t,\Delta\bm{x})&=\braket{0|\hat\phi(t,\bm{x})\hat\phi(t',\bm{x}')|0}\nonumber\\
    &= \int \frac{\d{n}{\bm k}}{2 (2\pi)^n\omega_{\bm{k}}} e^{-\ii\omega_{\bm{k}}\Delta t +\ii \bm{k}\cdot \Delta \bm{x}}.
\end{align}

For a state of the form~\eqref{eq:rho_qubits}, it is well known that we can quantify the amount of entanglement using the negativity $\mathcal N$~\cite{orignegativity1998,negativity2002}. For the final state $\hat\rho_\textsc{ab}$ given in Eq.~\eqref{eq:rho_qubits} the negativity is~\cite{Reznik2003, Pozas2015}:
\begin{equation}
    \mathcal N = \max(|\mathcal M| - \mathcal L, 0) + \mathcal O(\lambda^4).\label{eq:negativity}
\end{equation}

\section{Communication vs genuine harvesting}
\label{sec:communication}

In this section we will review the decomposition of the entanglement acquired by the detectors in a communication component and a genuine entanglement component first introduced in~\cite{Erickson2021_When}. 

For an arbitrary state of the field $\hat\rho_\phi$, the two-point correlator is defined as
\begin{equation}
    W(\mathsf{x},\mathsf{x}') = \braket{\hat\phi(\mathsf{x})\hat\phi(\mathsf{x}')}_{\hat\rho_\phi},
\end{equation}
which can be divided into symmetric and antisymmetric parts,
\begin{equation}
    W^{\pm}(\mathsf{x},\mathsf{x}')= \frac{W(\mathsf{x},\mathsf{x}')\pm W(\mathsf{x}',\mathsf{x})}{2}.
\end{equation}
Moreover, $W^+$ and $W^-$ respectively are the real and imaginary parts of $W$ due to $W(\mathsf{x}',\mathsf{x})=W^*(\mathsf{x},\mathsf{x}')$, and
\begin{align}
        W^+(\mathsf{x},\mathsf{x}')&=\frac{1}{2}\braket{\{\hat\phi(\mathsf{x}),\hat\phi(\mathsf{x}')\}}_{\hat\rho_\phi},\nonumber\\
        W^-(\mathsf{x},\mathsf{x}')&=\frac{1}{2}\braket{[\hat\phi(\mathsf{x}),\hat\phi(\mathsf{x}')]}_{\hat\rho_\phi}.
\end{align}
In~\cite{Erickson2021_When} it was argued that in a situation where two detectors get entangled through a quantum field while in causal contact, (hence the acquired entanglement could come both from harvesting and from communication), the contribution of $W^+$ is the one that one could associate with genuinely harvested entanglement. We summarize the evidence for that conclusion in the following: 
\begin{enumerate}
    \item The expectation value of the field commutator $[\hat\phi(\mathsf{x}),\hat\phi(\mathsf{x}')]$ is independent of the state, while the field anticommutator $\{\hat\phi(\mathsf{x}),\hat\phi(\mathsf{x}')\}$ does depend on the sate. Therefore, is independent of any preexisting correlations in the field, which are dictated by the field state. Any entanglement acquired between the detectors that is mediated by this term (without affecting the local noise) would still be the same even if one reduces the pre-existing entanglement in the field state, hence the state-independent contributions to entanglement coming from $W^-$ cannot be due to entanglement harvested form the field.
    
    \item The field commutator is precisely the difference of the retarded Green's function and the advanced one (the so-called causal propagator). Notice that if the quantum field theory were to be replaced by a classical field theory (where there is no entanglement) the role of the commutator would be played by the Poisson bracket and the communication through the classical field would be mediated by this very same Green function. These classical Green's functions are what mediate communication between interacting sources of the field and are zero outside the light cone. 

    Conversely, the field anticommutator has support even for spacelike separated events, and thus $W^+$ provides the only contribution to spacelike entanglement harvesting.

    \item It was shown in~\cite{cavitySignaling2014,infoWOenergy2015,detectorsSignaling2015,Pipo2023Signaling} that the communication between two detectors is mediated, at leading order, by the field commutator. Therefore, this points towards the entanglement gathered due to the contributions of $W^-$ at leading order to be communication-mediated. Meanwhile, the field anticommutator never contributes to communication at leading order, and thus, at leading order, one can view the contributions of $W^+$ to entanglement as genuine entanglement harvesting.

    More generally, two systems A, B that couple to the interaction picture observables $\hat C_\textsc{a}(t)$ and $\hat C_\textsc{b}(t')$ of a third system C, communicate at leading order solely through $[\hat C_\textsc{a}(t),\hat C_\textsc{b}(t')]$. Furthermore, even non-perturbatively, A and B cannot communicate if they couple to commuting observables of C in the interaction picture (see, e.g., Appendix \ref{apx:signaling}).
\end{enumerate}
For the negativity in Eq.~\eqref{eq:negativity}, the term $|\mathcal M|$ amounts to the correlations that the detectors gathered that contribute positively to the entanglement. Following the arguments above, we split this correlation term $\mathcal M$ by source: pre-existing field correlations $\mathcal M^+$ and communication $\mathcal M^-$. Concretely,
\begin{equation}
    \mathcal M = \mathcal M^+ + \mathcal M^-,
\end{equation}
where $\mathcal M^\pm$ only contains the contribution of $W^\pm$.

\section{Cancellation of communication in causal contact for derivative coupling}
\label{sec:results}

Now we move onto showing how the behaviour of the imaginary part of the Wightman function (and therefore of $\mathcal M^-$) is affected by coupling to $\partial_t \hat\phi$ instead of $\hat\phi$.  Remarkably, we can find regimes in Minkowski spacetimes where the communication-mediated correlations $\mathcal M^-$ can become zero in full light-like contact while the entanglement gathered by the detectors is maximum.

Denoting as $W_n^-$ the antisymmetric part of the two-point function of the vacuum of ($n$+1)-dimensional Minkowski spacetime (for a derivation of the expressions below see, e.g., \cite{Erickson2021_When}),
\begin{equation}
    W_1^-(\mathsf{x},\mathsf{x}')=-\frac{\ii }{4} \big(\Heaviside(|\Delta x|+\Delta t)-\Heaviside(|\Delta x|-\Delta t)\big),
\end{equation}
\begin{equation}
    W_2^-(\mathsf{x},\mathsf{x}')=-\frac{\ii \sign(\Delta t)}{4 \pi}\frac{ \Heaviside\left(\Delta t^2-\Delta \bm{x}^2\right)}{\sqrt{\Delta t^2-\Delta \bm{x}^2}},
\end{equation}
and, for odd $n\geq3$,
\begin{align}
    W_n^-(\mathsf{x},\mathsf{x}')&=\sum_{j=0}^{\frac{n-3}{2}}\frac{\ii a_j}{|\Delta \bm{x}|^{n-2-j}}\big(\delta^{(j)}(\Delta t+|\Delta \bm{x}|)\nonumber\\
    &\qquad-(-1)^{j}\delta^{(j)}(\Delta t-|\Delta \bm{x}|)\big),
\end{align}
where $a_j$ are real numbers and $\delta^{(j)}$ are the Dirac delta distribution's $j$-th derivative. 

The two time derivatives of $W_n^-$ that appear in $\mathcal M^-$ in Eq.~\eqref{eq:LMSimple} are the only change introduced by coupling to $\partial_t\hat\phi$ instead of $\hat\phi$. Therefore, for amplitude coupling, as discussed in the section above and in~\cite{Erickson2021_When}, communication is mediated by $W_n^-$, while, for derivative coupling, the communication is mediated by $\partial_t\partial_{t'} W_n^-$. 

For $n=1$,
\begin{align}
    \partial_t\partial_{t'}W_1^-(\mathsf{x},\mathsf{x}')=\frac{\ii }{4} \big(\delta^{(1)}(\Delta t+|\Delta x|)+\delta^{(1)}(\Delta t-|\Delta x|)\big)\label{eq:W1-}.
\end{align}

For $n=2$, the following closed expression is valid for any $\mathsf{x}$, $\mathsf{x}'$ that are not light-connected:
\begin{align}
    &\partial_t\partial_{t'}W_2^-(\mathsf{x},\mathsf{x}')\nonumber\\
    &=\frac{\ii\sign(\Delta t)}{4 \pi}\frac{(2 \Delta t^2+\Delta \bm{x}^2)\Heaviside(\Delta t^2-\Delta \bm{x}^2)}{\left(\Delta t^2-\Delta \bm{x}^2\right)^{5/2}},
\end{align}
and, in any case, for any $\mathsf{x}$, $\mathsf{x}'$ the distribution can be evaluated from the following integral form:
\begin{align}
     &\partial_t\partial_{t'}W_2^-(\mathsf{x},\mathsf{x}')=-\frac{\ii}{4 \pi }\int_0^\infty\diff \omega \omega^2 \sin (\omega \Delta t ) J_0(\omega  |\Delta \bm{x}| ),
\end{align}
where $J_0$ is the Bessel function of the first kind of order zero.

Furthermore, for odd $n\geq3$,
\begin{align}
    \partial_t\partial_{t'}W_n^-(\mathsf{x},\mathsf{x}')
    &= \sum_{j=0}^{\frac{n-3}{2}}\frac{-\ii a_j}{|\Delta \bm{x}|^{n-2-j}}\big(\delta^{(j+2)}(\Delta t+|\Delta \bm{x}|)\nonumber\\
    &\qquad-(-1)^{j}\delta^{(j+2)}(\Delta t-|\Delta \bm{x}|)\big).
\end{align}
Here, notice that for large enough distances $|\Delta\bm{x}|$, the term with \mbox{$j=\frac{n-3}{2}$} dominates, decaying as $|\Delta\bm{x}|^{-\frac{n-1}{2}}$. Interestingly, in this long-distance regime, at least for $n$ odd, $\partial_t\partial_{t}W^-_n$ has a behaviour akin to $W^-_{n+4}$, except that it keeps the $|\Delta \bm{x}|$ decay of $W^-_n$. 

These expressions show that both $W_n^-$ and $\partial_t\partial_{t'}W_n^-$ are only supported in the light-cone (and inside the light-cone for $W_1^-$, $W_2^-$ and $\partial_t\partial_{t'}W_2^-$), so one might expect that $\mathcal M^-$ peaks at the light-cone. This is the case when coupling to the amplitude in $1+1$ and $3+1$ dimensions, with, for example, Gaussian switching functions \cite{Erickson2021_When}. However, this changes when we couple to the derivative. Then, $\mathcal M^-$ can reach zero even in full light contact (in $1+1$ D) or in partial light contact (in $3+1$ D), as we show next. 

For simplicity, consider Gaussian profiles for both centered switching and smearing functions (see Eq.~\eqref{eq:translation}),
\begin{equation}
    \chi(t)=e^{-\frac{t^2}{T^2}},\ F(\bm{x})=\frac{1}{(\sqrt{\pi}\sigma)^n}e^{-\frac{\bm{x}^2}{\sigma^2}},\label{eq:switching_smearing}
\end{equation}
where $T$ is the interaction time-scale, and $\sigma$ is the size of the detector. Then, from Eq.~\eqref{eq:LMSimple},
\begin{align}
    &\mathcal M=CT\!\int\! \diff{t_-}\d{n}{\bm{x}_-}\!   \frac{e^{-\frac{t_-^2}{2T^2}-\frac{\bm{x}_-^2}{2\sigma^2}}}{(\sqrt{2\pi}\sigma)^n}W_{tt'}(|t_-\! -t_\Delta|,\bm{x}_-\! -\bm{x}_\Delta),\nonumber\\
    &C= -\sqrt{\frac{\pi}{2}}\lambda^2e^{\ii\Omega(t_\textsc{a}+t_\textsc{b})-\frac{1}{2}(\Omega T)^2},
\end{align}
where we took the change of variables $t_\pm = t\pm t'$, \mbox{$\bm{x}_\pm = \bm{x}\pm \bm{x}'$} and then integrated over $t_+$ and $\bm{x}_+$. 
\begin{figure*}[ht]
    \centering
    \includegraphics[width=1.75\columnwidth]{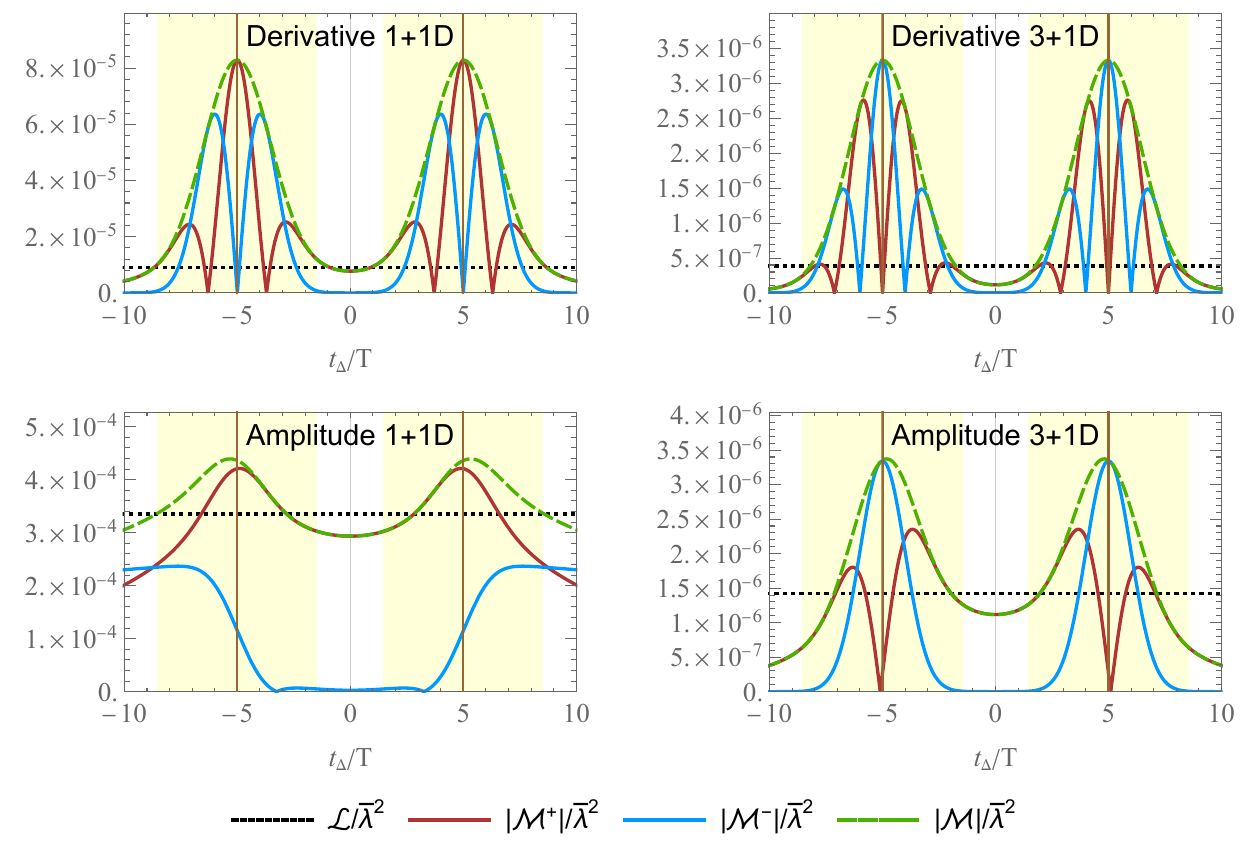}
    \caption{Contributions to the negativity $\mathcal N$, see Eq.~\eqref{eq:negativity}, for a series of derivative coupling and amplitude coupling scenarios. The correlations $|\mathcal M|$ which contribute to $\mathcal N$ are split between the $|\mathcal M^+|$ term (genuine harvesting)  the $|\mathcal M^-|$ term (acquired by communication). The setup uses the Gaussian switching and smearing functions given in Eq.~\eqref{eq:switching_smearing},\mbox{ $\Omega T= 4$, $|x_\Delta|/T=5$, $\sigma/T=0.05$}. The amplitude coupling in 1+1D has an IR cutoff of $\Lambda T = 0.02$, see, e.g. \cite{Pozas2015}. The $\overline{\lambda}$ is an adimensional coupling constant, whose definition depends on the scenario: for \textit{derivative coupling}, $\overline{\lambda}=\lambda$ in $1+1$D and $\overline{\lambda}=\lambda/T$ in 3+1D. For \textit{amplitude coupling}, $\overline{\lambda}=\lambda T$ in $1+1$D and $\overline{\lambda}=\lambda$ in 3+1D.
    The brown vertical lines denote maximum light contact, and the light yellow coloured regions show the size of the strong support of the switching function of a detector, taken to be $[-3.5T,3.5T]$.  }
    \label{fig:derivCouplingComb}
\end{figure*}
To better understand how $\mathcal M^-$ for $n=1$ becomes zero in full causal contact, we integrate over $t_-$ analytically. Using that $W_{tt'}= \partial_t\partial_{t'} W$ together with Eq.~\eqref{eq:W1-},
\begin{align}
    &\mathcal M^-_1=\frac{\ii C}{4}\int \diff{x_-}    \bigg(\frac{|x_- -x_\Delta|+t_\Delta}{T}e^{-\frac{1}{2}\big(\frac{|x_--x_\Delta|+t_\Delta}{T}\big)^2}\nonumber\\
    &\qquad+\frac{|x_- -x_\Delta|-t_\Delta}{T}e^{-\frac{1}{2}\big(\frac{|x_--x_\Delta|-t_\Delta}{T}\big)^2}\bigg)\frac{e^{-\frac{x_-^2}{2\sigma^2}}}{\sqrt{2\pi}\sigma},
\end{align}
which becomes easier to interpret in the point-like detector limit, where $\sigma \to 0$, and for $(|x_\Delta|+|t_\Delta|)\gg T$, so that the causal relationships are well-defined. In these regimes,
\begin{align}
    &\mathcal M^-_1\approx\frac{\ii C}{4}\frac{|x_\Delta|-|t_\Delta|}{T}e^{-\frac{1}{2}\big(\frac{|x_\Delta|-|t_\Delta|}{T}\big)^2}.
\end{align}
Therefore, for derivative coupling in $1+1$ dimensions, $|\mathcal M^-|$ is gaussianly suppressed as we move away from the light-cone, but actually cancels on the light-cone itself (where $|x_\Delta|=|t_\Delta|$), as seen in Figure~\ref{fig:derivCouplingComb}(a). Simultaneously, $|\mathcal M^+|$ and $|\mathcal M|$ reach a maximum at maximum light-like communication, causing the detectors to gather the most entanglement, and this entanglement being genuinely harvested from the field.

For derivative coupling in $3+1$ dimensions, $\mathcal M^-$ results from integrating a Gaussian against $\delta^{(2)}$ (in the pointlike limit $\sigma\to0$), which causes $|\mathcal M^-|$ to have three peaks, as seen in Figure~\ref{fig:derivCouplingComb}(b), with communication reaching zero in between these peaks, even though the detectors are in causal contact. This phenomenon allows to genuinely harvest entanglement from a massless field also in $3+1$ dimensions even in partial light-like contact, contrarily to the amplitude coupling case \cite{Erickson2021_When}.

\section{Conclusions}

When a pair of particle detectors in causal contact get entangled through their interaction with a quantum field, the entanglement between the detectors could generally come from 1) their exchange of information through the field, and 2) from from extracting pre-existing entanglement in the field. When the second contribution dominates we can say that the detectors are \textit{harvesting} entanglement from the field. In previous results it was shown than in most common cases in flat spacetime, when the (the interaction regions of the) detectors are in full light contact, the entanglement they acquire is dominated by communication~\cite{Erickson2021_When} (with interesting exceptions to this rule appearing in spacetimes with caustics~\cite{Lensing2023}).  

In contrast, here we showed that if the particle detectors couple through a massless field in flat spacetimes through the so-called \textit{derivative coupling}, it is possible to have the peak of entanglement between the detectors happening while they are in full light contact, and (more importantly) the source of that entanglement is genuine harvesting and not communication.

This is relevant since the derivative coupling is arguably important both from the theoretical point of view (to sidestep IR ambiguity problems and to study problems in reduced dimensions) but also is relevant in the modelling of superconducting qubits coupled to transmission lines.

\acknowledgements

A. T.-B. received the support of a fellowship from ``la Caixa” Foundation (ID 100010434, with fellowship code LCF/BQ/EU21/11890119). Research at Perimeter Institute is supported in part by the Government of Canada through the Department of Innovation, Science and Industry Canada and by the Province of Ontario through the Ministry of Colleges and Universities. EMM acknowledges support through the Discovery Grant Program of the Natural Sciences and Engineering Research Council of Canada (NSERC). EMM  thanks the support from his Ontario Early Researcher award.

\appendix
\section{Communication at leading order through a mediating system}
\label{apx:signaling}
Here we provide an quick review on how to compute estimator for the ability of a system A to send a signal to a system B, through a mediating system C, at leading perturbative order. This scenario generalizes the case of a pair of detectors communicating through a quantum field. For a more thorough analysis see~\cite{detectorsSignaling2015,Pipo2023Signaling}. 

To quantify communication from A to B, it is useful to distinguish between the full time evolution $\hat U$, and the time evolution if we only allow the system $\nu\in\{\text{A},\text{B}\}$ and C to interact, which we denote as $\hat U_\nu$. Then, all the conributions to the final state of B coming from the existence of  A, i.e. how A influences the final state of B, can be expressed as
\begin{equation}
    \hat\rho_{\textsc{b}}^\text{signal} = \Tr_\textsc{ac}(\hat U \hat\rho_0\hat U^\dagger - \hat U_\textsc{b} \hat\rho_0\hat U_\textsc{b}^\dagger),\label{eq:def_com}
\end{equation}
for any initial state $\hat\rho_0$.

For simplicity, we restrict ourselves to interaction Hamiltonians (in the interaction picture) of the form
\begin{align}
    &\hat{H}_I(t)=\hat{H}_\textsc{a}(t) + \hat{H}_\textsc{b}(t),\nonumber\\
    &\hat{H}_\nu(t) = \hat O_\nu(t) \hat C_\nu(t).\label{eq:HAHB}
\end{align}
Here, $\hat O_\textsc{a}(t)$ and $\hat O_\textsc{b}(t)$ are observables of the system A and B respectively. The operators $\hat C_\textsc{a}(t)$ and $\hat C_\textsc{b}(t)$ are observables of the mediating system C.

To further showcase the role of $[\hat C_\textsc{a}(t),\hat C_\textsc{b}(t')]$ in communication we explicitly compute $\hat\rho_{\textsc{b}}^\text{signal}$ to the leading order in a perturbative expansion. Using the Dyson series of Eq.~\eqref{eq:DysonSeries},\begin{align}
     \hat\rho_{\textsc{b}}^\text{signal} &= \Tr_\textsc{ac}\Bigl(\hat U_\textsc{b}^{(1)} \hat\rho_0 \hat U_\textsc{a}^{\dagger(1)}+\big(\hat U_\textsc{ab}^{(2)} +\hat U_\textsc{ba}^{(2)} \big)\hat\rho_0 + \text{H.c.}\Bigr) \nonumber\\
    &\quad +\mathcal{O}\bigl(\hat O_\textsc{a}^{m} \hat O_\textsc{b}^{l}\bigr),
\end{align}
where $m,l\geq 1$, $m+l\geq 3$ and we defined
\begin{align}
    &\hat U_\nu^{(1)}=-\ii\int\diff {t} \hat H_\nu(t),\nonumber\\
    &\hat U_{\mu\nu}^{(2)}=-\int\diff {t}\int^t \diff {t'} \hat H_\mu(t)\hat H_\nu(t').
\end{align}
Then, using the properties of the trace, and that observables from different systems commute,
\begin{align}
    &\Tr_\textsc{ac}\bigl(\hat U_\textsc{b}^{(1)} \hat\rho_0 \hat U_\textsc{a}^{\dagger(1)}\bigr)\nonumber \\
    &= \int\diff {t}\diff {t'} \hat O_\textsc{b}(t)\Tr_\textsc{ac}\bigl( \hat\rho_0 \hat O_\textsc{a}(t')  \hat C_\textsc{a}(t')\hat C_\textsc{b}(t)\bigr),
\end{align}

\begin{align}
    &\Tr_\textsc{ac}\bigl((\hat U_\textsc{ab}^{(2)} +\hat U_\textsc{ba}^{(2)})\hat\rho_0 \bigr)\nonumber \\
    &= -\int\diff {t} \diff {t'} \hat O_\textsc{b}(t)\Tr_\textsc{ac}\Bigl(  \hat\rho_0 \hat O_\textsc{a}(t')\nonumber \\
    &\qquad \cross \big(\Heaviside(t'-t)\hat C_\textsc{a}(t') \hat C_\textsc{b}(t) +\Heaviside(t-t')\hat C_\textsc{b}(t) \hat C_\textsc{a}(t')\big)\Bigr),
\end{align}
where $\Heaviside$ is the Heaviside step function. Substituting back into $\hat\rho_{\textsc{b}}^\text{signal}$,
\begin{align}
    \hat\rho_{\textsc{b}}^\text{signal}& = \int\diff {t}\int_t\diff {t'} [\hat O_\textsc{b}(t'),\Tr_\textsc{ac}\bigl( \hat\rho_0 \hat O_\textsc{a}(t)  [\hat C_\textsc{a}(t),\hat C_\textsc{b}(t')]\bigr)]\nonumber\\
    &\qquad+\mathcal{O}\bigl(\hat O_\textsc{a}^{m} \hat O_\textsc{b}^{l}\bigr),
\end{align}
with $m,l\geq 1$, $m+l\geq 3$. Therefore, at leading order, the communication is mediated solely by the commutator $[\hat C_\textsc{a}(t),\hat C_\textsc{b}(t')]$, as we wanted to show. 
Notice that $[\hat C_\textsc{a}(t),\hat C_\textsc{b}(t')]$ only contributes for $t'\geq t$, i.e. A can only message B through $\hat C_\textsc{b}(t')$ after A has coupled to $\hat C_\textsc{a}(t)$, as would be expected. This order adequately reverses if we consider communication from B to A.

\emph{Non-perturbative case:} Remarkably, communication is impossible if the observables of system C in Eq.~\eqref{eq:HAHB} commute, i.e. if \mbox{$[\hat C_\textsc{a}(t),\hat C_\textsc{b}(t')]=0$}, then \mbox{$\hat\rho_{\textsc{b}}^\text{signal}=0$}. This follows from noticing that, under this assumption, \mbox{$[\hat H_\textsc{a}(t),\hat H_\textsc{b}(t')]=0$}, and thus $\hat U = \hat U_\textsc{a}\hat U_\textsc{b}=\hat U_\textsc{b}\hat U_\textsc{a}$, which after substituting in Eq.~\eqref{eq:def_com} and using the cyclic property of the trace together with $\hat U_\textsc{a}\hat U_\textsc{a}^\dagger=\id$ shows that $\hat\rho_{\textsc{b}}^\text{signal}=0$. In other words, communicating through C requires coupling to non-commuting interaction picture observables of C. Notice that the same is not necessarily true for gaining correlations. A and B can get correlated even if they couple to commuting interaction picture observables of C, by extracting pre-existing correlations in C.

\bibliographystyle{apsrev4-2}
\bibliography{references}
\end{document}